\documentclass[prd,nofootinbib,showpacs,twocolumn]{revtex4-1}
\usepackage{color}
\usepackage{bm}
\usepackage{graphicx}
\usepackage{amsmath}
\usepackage{amssymb}
\usepackage{enumitem}
\usepackage{hyperref}
\usepackage{subfigure}
\usepackage{array}
\usepackage[english]{babel}
\usepackage{tensor}
\usepackage{comment}
\usepackage[normalem]{ulem}
\usepackage[dvipsnames]{xcolor}

\def\be{\begin{equation}}
\def\ee{\end{equation}}
\def\ba{\begin{eqnarray}}
\def\ea{\end{eqnarray}}
\def\l{\left}
\def\r{\right}
\def\f{\frac}

\allowdisplaybreaks

\def\nn{\nonumber}

\def\hub{{\mathcal H}}

\begin{document}

\title{Do current cosmological observations rule out all Covariant Galileons?}

\author{Simone Peirone$^1$, Noemi Frusciante$^2$, Bin Hu$^3$, Marco Raveri$^{4,1}$, Alessandra Silvestri$^1$}
\affiliation{
\smallskip
$^{1}$  Institute Lorentz, Leiden University, PO Box 9506, Leiden 2300 RA, The Netherlands \\
\smallskip
$^{2}$  Instituto de Astrofisica e Ci\^{e}ncias do Espa\c{c}o, Faculdade de Ci\^{e}ncias da Universidade de Lisboa, Edificio C8, Campo Grande, P-1749016, Lisboa, Portugal \\  
\smallskip
$^{3}$  Department of Astronomy, Beijing Normal University, Beijing, 100875, China \\
\smallskip 
$^{4}$  Kavli Institute for Cosmological Physics, Enrico Fermi Institute,The University of Chicago, Chicago, Illinois 60637, USA}

\begin{abstract}
We revisit the cosmology of Covariant Galileon gravity in view of the most recent cosmological data sets, including weak lensing. 
As a higher derivative theory, Covariant Galileon models do not have a $\Lambda$CDM limit and predict a very different structure formation pattern compared with the standard $\Lambda$CDM scenario.  
Previous cosmological analyses suggest that this model is marginally disfavoured, yet can not be completely ruled out.  In this work we use a more recent and extended combination of data, and we allow for more freedom in the cosmology, by including a massive neutrino sector with three different mass hierarchies.
We use  the Planck measurements of Cosmic Microwave Background temperature and polarization;  Baryonic Acoustic Oscillations measurements by BOSS DR12; local measurements of $H_0$; the joint light-curve analysis supernovae sample; and, for the first time, weak gravitational lensing from the KiDS collaboration.  
We find,  that in  order to provide a reasonable fit, a non-zero neutrino mass is indeed necessary, but we do not report any sizable difference among the three neutrino hierarchies.
Finally, the comparison of the Bayesian Evidence to the $\Lambda$CDM one  shows that in all the cases considered, Covariant Galileon models are statistically ruled out by cosmological data.
\end{abstract}


\date{\today}

\maketitle
\section{Introduction}\label{Intro}

 Covariant Galileon (CG) models~\cite{Deffayet:2009wt} are a class of scalar-tensor theories belonging to the broader class of Generalized Galileons, i.e. scalar-tensor theories with second order equations of motion~\cite{Horndeski:1974wa,Deffayet:2009mn}. Galileon theories  gained interest in the past years because they allow for self accelerating solutions that could describe both the inflationary epoch  and the late time accelerated expansion~\cite{Kobayashi:2010cm,Burrage:2010cu,Kobayashi:2011nu,Deffayet:2010qz}. 

Along with a modification of the background expansion history, CG models lead to peculiar features in the large scale structure~\cite{Barreira:2013xea,2014JCAP...08..059B,Renk:2017rzu}, in particular contributing in enhancing the low-$\ell$ part of the Cosmic Microwave Background (CMB) lensing spectrum. 
Quartic and Quintic models (namely those including terms up to quartic and quintic order in the scalar field, respectively), are preferred by Planck data because they predict a  lower impact of the integrated Sachs-Wolfe (ISW) effect, but at the same time it is hard for them to pass the Solar System constraints, which are better accommodated by the Cubic model (as previously, this model contains three copies of the scalar field).  Moreover, as shown in~\cite{2014JCAP...08..059B}, the CG model prefers nonzero neutrino masses at over 5~$\sigma$, which in turns affect the $H_0$ estimation, making it compatible with local measurements.

In this paper we analyze the cosmology of CG models in light of the current cosmological observations, including, for the first time, data from the weak gravitational lensing (WL) survey KiDS~\cite{Hildebrandt:2016iqg,Kuijken:2015vca,deJong:2012zb}. Previous works~\cite{Barreira:2013xea,2014JCAP...08..059B,Renk:2017rzu} have shown that for the CG model it is hard to provide a fit to data better than the standard cosmological model and  a recent analysis showed indeed that the Cubic branch can be ruled out  at 7.8$\sigma$  with data including CMB, BAO and ISW. However the Quartic and Quintic models can not be completely excluded by such collection of data.
In this work, we  extend the analysis by using more recent data sets and adding the WL measurements, simultaneously allowing for different mass hierarchies in the massive neutrinos sector. 
 
The cosmological impact of the mass hierarchy has not been explored extensively. In general, it is expected that the sensitivity to the type of hierarchy increases as the bound on the total mass of neutrinos becomes tighter, see e.g.~\cite{Hannestad:2016fog}. Only very recently it has been shown that in the $\Lambda$CDM scenario there is a mild preference for the normal hierarchy~\cite{Hannestad:2016fog,Simpson:2017qvj,Vagnozzi:2017ovm} and that, in models with a parametrized dark energy equation of state, different hierarchies seem to have a slight impact on the dark energy parameters while leaving unaffected the standard cosmological  parameters~\cite{2017PhRvD..95j3522Y}. Additionally, the different hierarchies imply different transition redshifts from relativistic to non-relativistic regimes, and this would leave an impact on the matter power spectrum. Such an effect has been usually neglected, because it is negligible if compared to the effect coming from the the total neutrino mass. However, data coming from last generation of surveys and  future experiments, such as EUCLID,  might have the accuracy needed to constraint these features~\cite{DeBernardis:2009di}. 

Furthermore, the inclusion of new parameters, let them be the parameters specific of the CG model or to the neutrinos sector, could allow to ease  the tensions between CMB measurements and low redshift data, concerning both the local measurements of the Hubble constant as well as WL measurements. The first year results from DES collaboration~\cite{Abbott:2017wau,Abbott:2017smn} show that the CMB-WL tension on $S_8 = \sigma_8\sqrt{\Omega_m/0.3}$ is somehow reduced from $2.3 \, \sigma$ (KiDS vs Planck) to $1.6 \, \sigma$ (DES vs Planck). Similarly, also the CMB-$H_0$ tension seems to be reduced by DES measurements, relying on the large  error bars of the first year release. While these are barely statistically significant, and likely to be settled, it is still interesting to investigate  them within the framework of extended models such as the CG.
The use of WL data, a novel aspect of our analysis, has a relevant role in studying the CMB-WL tension.

Finally,  the recent multi-messenger observation of the binary neutron star merger~\cite{TheLIGOScientific:2017qsa,Coulter:2017wya,GBM:2017lvd} was shown to cast stringent constraints on the Quartic and Quintic Galileon Lagrangians, practically ruling them out as dark energy candidates~\cite{Creminelli:2017sry,Ezquiaga:2017ekz,2017arXiv171006394B,Sakstein:2017xjx, Bettoni:2016mij}.  In this work we use  a complementary and entirely independent set of data, from cosmological observations, and derive very stringent constraints on all three CG Lagrangians.

The manuscript is organized as follows. In Section~\ref{model}, we review the CG model with its background evolution when a tracker solution is considered and we summarize the definitions of the neutrinos hierarchies. In Section~\ref{method}, we introduce the Einstein Boltzmann code and the data sets used for the analysis. In Section~\ref{results}, we discuss the results and in Section~\ref{conclusion}, we draw our conclusions.

\section{The model} \label{model}
\subsection{Covariant Galileons}
The Galilean symmetry $\partial_\mu \phi \rightarrow \partial_\mu\phi+b_\mu$ (being $b_\mu$ a constant) has been considered to construct the most general action with a metric ($g_{\mu\nu}$) and a scalar field ($\phi$),  whose field equations include up to second order derivatives~\cite{Nicolis:2008in}, thus avoiding Ostrogradski instabilities~\cite{Ostrogradsky:1850fid}. The first formulation was in flat space and the generic structure of the Galileon Lagrangian terms follows  $\partial \phi \cdot \partial\phi (\partial^2\phi)^{n-2}$, up to $n = 5$, as higher order Lagrangians are just total derivatives.  The same approach has been generalized on a curved space-time, but in this case, in order to retain second order field equations and ensure the propagation of only one additional degree of freedom, extra terms non minimally coupled to the metric have been added to the action~\cite{Deffayet:2009wt}. This ended up with the loss of the Galileon symmetry, while preserving the shift symmetry. The resulting model is known as CG and the action reads
\ba\label{CGaction}
S_{CG}&=&\int{}d^4x\sqrt{-g}\l\{\f{m_0^2}{2}R-\frac{1}{2}c_2 X+\frac{c_3}{M^3} X \Box\phi \r.\nn\\
&&+\l. \frac{c_4}{4 M^6} X^2 R -\frac{c_4}{ M^6}X\left[ (\Box\phi)^2 - \phi^{;\mu \nu} \phi_{;\mu \nu}\right] \r.\nn\\
&&+\l.\frac{3 c_5}{4 M^9} X^2 G_{\mu \nu}\phi^{;\mu \nu}+\frac{ c_5}{2 M^9}X\left[ (\Box\phi)^3 \r.\r.\nn\\
&&\l.\l.-3\Box\phi \,  \phi^{;\mu \nu} \phi_{;\mu \nu}+2   \phi^{;\mu \nu} \phi_{;\mu \sigma}\phi_{;\sigma}^{;\nu}\right]\r\}\,,
\ea
where $m_0^2$ is the Planck mass, $g$ is the determinant of the metric, $R$ and $G_{\mu\nu}$ are respectively the Ricci scalar and the Einstein tensor,  $X=\phi^{;\mu}\phi_{;\mu}$ is called kinetic term and $	\{;\}$ stands for the covariant derivative. Moreover, $c_i$ are constant dimensionless parameters and $M^3=m_0H_0^2$ with $H_0$ being the present time value of the Hubble parameter. 

In order to investigate the reliability of this model on cosmological linear scales, we will exploit the tracker solution for the background evolution~\cite{DeFelice:2010pv}. The tracker solution relates the scalar field and the Hubble parameter as follows
\be
\left(\frac{\hub}{a}\right)^2 \psi=\xi H_0^2=const, \label{tracker}
\ee
where $\psi=\f{1}{m_0}\frac{d \phi}{d \ln a}$ is a dimensionless field,  $\xi$ is a dimensionless constant  and $\hub\equiv \f{da}{ad\tau}$ is the conformal Hubble parameter.  Then, Eq.~\ref{tracker} can be used  to obtain the expansion history, $\hub$ along the tracker~\cite{2014JCAP...08..059B}. Indeed, assuming the tracker solution and a flat Friedmann-Lemaitre-Robertson-Walker metric with signature ($-,+,+,+$), the modified Friedmann equation can be written as follows:
\ba
E^4(a)&=&E^2(a) \left( \Omega_{m,0} a^{-3} +\Omega_{r,0} a^{-4} + \Omega_{\nu,0} \frac{\rho_\nu(a)}{\rho_{\nu,0}}\right) \nn\\
&+& \left[ \frac{c_2}{6 } \xi^2 + 2 c_3 \xi^3+ \frac{15}{2}c_4 \xi^4 +7 c_5 \xi^5\right]\,, \label{equation_E}
\ea
where $E=\f{\hub}{aH_0}$ and $\Omega_{i,0}$ stand for the present density parameters for baryons and cold dark matter ($m=b,cdm$), radiation and massless neutrinos (r) and massive neutrinos ($\nu$). Then, Eq.~\eqref{equation_E} can be solved to get $\hub$. Along with this equation, one has to consider two further constraints: one comes from the flatness condition, which immediately gives the definition of the present density parameter for the scalar field:
\be
\Omega_{\phi,0} = \frac{c_2}{6 } \xi^2 + 2 c_3 \xi^3 + \frac{15}{2}c_4 \xi^4 +7 c_5 \xi^5 \,, \label{flatness}
\ee
the second is obtained combining  the equation for the scalar field (obtained by varying the action with respect to the scalar field) and Eq.~\eqref{tracker}, which gives
\be 
c_2 \xi + 6 c_3  \xi^2 +18 c_4 \xi^3 + 15c_5 \xi^4 =0 \,. \label{second_constraint}
\ee
Finally, to avoid scaling degeneracy one has the freedom to fix $c_2=-1$ without loss of generality. 

In the present work, we will analyze three sub-classes of CG model and the constraints~\eqref{flatness}-\eqref{second_constraint} will be used to  define the corresponding sets of free parameters as follows:
\begin{itemize}
\item $G_3$: Cubic model,  $c_3\neq 0\,,\{c_4,c_5\}=0$.\\
Using the constraint relations, one has
\ba 
\xi = \sqrt{6 \Omega_{\phi,0}}\,,\quad c_3 =\frac{1}{6  \sqrt{6 \Omega_{\phi, 0}}}\,.
\ea
Thus, in this case the number of free parameters is the same as in $\Lambda$CDM.
\item $G_4$: Quartic model, $\{c_3\,,c_4\}\neq 0\,, c_5=0$.
We have:
\ba
c_3&=&\frac{1}{2} \xi^{-1} - 2 \Omega_{\phi, 0} \xi^{-3}\,,\\
c_4&=&-\frac{1}{9} \xi^{-2} +\frac{ 2}{3} \Omega_{\phi, 0} \xi^{-4}\,,
\ea
with one extra free parameter $\xi$.
\item $G_5$: Quintic model, $\{c_3,c_4,c_5\}\neq 0$.

\noindent
Solving Eqs.~\eqref{flatness}-\eqref{second_constraint} for $c_4$ and $c_5$, one gets
\ba 
c_5 &=& \frac{4}{3} \Omega_{\phi, 0} \xi^{-5} +\frac{1}{3} c_2 \xi ^{-3} +\frac{2}{3} c_3 \xi ^{-2} ,\nn\\
c_4 &=&-\frac{10}{9} \Omega_{\phi, 0} \xi^{-4} -\frac{1}{3} c_2 \xi ^{-2} -\frac{8}{9} c_3 \xi ^{-1}.
\ea
Then, one has $\{\xi, c_3\}$ as extra free parameters.
\end{itemize}

\subsection{Mass hierarchies}

It is well known that the mass of neutrinos leaves clear signatures on cosmological observables~\cite{2010gfe..book.....M}, such as a modification of the time at which matter-radiation  equality occurs, causing shifts of the first peak of the CMB temperature and polarization power spectra, via the early integrated Sachs-Wolfe (eISW) effect;  massive neutrinos also cause the free streaming of density perturbations on small scales, while behaving like clustering cold dark matter on larger scales. Cosmological analysis have established robust upper limits on the sum of the neutrino masses of $\Sigma m_\nu <0.13$~eV~\cite{Cuesta:2015iho}, $\Sigma m_\nu <0.12$~eV~\cite{Palanque-Delabrouille:2015pga} and very recently $\Sigma m_\nu <0.20$~eV~\cite{Abbott:2017smn} at $95 \%$ confidence level. 
Measurements of neutrino flavour oscillations imply that at least two neutrino species have nonzero masses~\cite{Gonzalez-Garcia:2014bfa} and the differences of the square of the neutrino masses are $\Delta_{12}^2 = m_2^2 -m_1^2 = 7.5 \times10^{-5} $ eV$^2$, from which it follows  $m_2>m_1$ and $|\Delta_{31}^2|=|m_3^2 -m_1^2| = 2.5\times10^{-3}$ eV$^2$, with $m_i$ being the mass of the $i$-th massive eigenstate. Since such experiments can just measure their differences, we are left with three possible mass hierarchies:  {\it normal hierarchy}, when  $m_3$ is taken to be the largest mass $m_3 \gg m_2 > m_1$; {\it inverted hierarchy}, $m_3$ is considered the smallest $m_2 > m_1\gg m_3$; the final option, the {\it degenerate hierarchy}, consists in considering that each mass is orders of magnitude bigger than each mass splitting ($m_j \sim m_i >> \Delta_{ij}$), thus  all three species are  treated as having effectively the same mass $m_1 = m_2 = m_3$. 

In the present analysis we allow for different mass hierarchies. This will permit to investigate possible effects due to the different free streaming length scales associated to the three neutrinos masses and see if the current cosmological data have the sensitivity to capture this effect. Moreover, the additional freedom connected to the choice of the hierarchy can be essential in order to make the CG models? predictions compatible with data.

\section{Method}\label{method}
\subsection{EFTCAMB}

We perform the present analysis by making use of  EFTCAMB/EFTCosmoMC\footnote{EFTCAMB webpage: \textcolor{blue}{\url{http://www.eftcamb.org}}}~\cite{Hu:2013twa,Raveri:2014cka}. These patches have been obtained by implementing the effective field theory approach for dark energy and modified gravity (hereafter EFT)~\cite{Gubitosi:2012hu,Bloomfield:2012ff,Gleyzes:2013ooa,Bloomfield:2013efa,Piazza:2013coa,Frusciante:2013zop,Gleyzes:2014rba} into CAMB/CosmoMC~\cite{Lewis:1999bs,Lewis:2002ah}.

In order to implement a specific model in EFTCAMB one has to implement the background evolution and provide a mapping between the free EFT functions $\{\Omega(a),\gamma_i(a)\}$, with $i=1...6$, and the model~\cite{Gubitosi:2012hu,Bloomfield:2012ff,Bloomfield:2013efa,Gleyzes:2013ooa,Gleyzes:2014rba,Frusciante:2015maa,Frusciante:2016xoj}. In the case under analysis, we have implemented the background evolution as in Eq.~\eqref{equation_E} and worked out the mapping as follows: 
\begin{eqnarray}
\Omega&=& \frac{a^4}{2 \mathcal{H}^4} H_0^4 \xi ^4 \left[c_4-6 c_5 \xi \left(1- \f{\dot{\mathcal{H}}}{\mathcal{H}^2}\right)\right]\,, \nn\\
\gamma_1 &=&\frac{a^2 H_0^2 \xi ^3}{4 \mathcal{H}^2}\Bigg[ 2c_4\xi\l(24 -\f{\ddot{\mathcal{H}}}{\mathcal{H}^3}-9\f{\dot{\mathcal{H}}}{\mathcal{H}^2}+5\f{\dot{\mathcal{H}}^2}{\mathcal{H}^4}\r)\nn\\
	&&+3c_5\xi^2\l(12+ 10\f{\ddot{\mathcal{H}}}{\mathcal{H}^3}+21 \f{\dot{\mathcal{H}}}{\mathcal{H}^2} +  \f{\dddot{\mathcal{H}}}{\mathcal{H}^4}-50\f{\dot{\mathcal{H}}^2}{\mathcal{H}^4}\r.  \nn\\ 
	&&\l.+42  \f{\dot{\mathcal{H}}^3}{\mathcal{H}^6}-18  \f{\dot{\mathcal{H}}}{\mathcal{H}^2}  \f{\ddot{\mathcal{H}}}{\mathcal{H}^3}\r)+2c_3\l(4-\f{\dot{\mathcal{H}}}{\mathcal{H}^2}\r)\Bigg]\,, \nn\\
\gamma_2 &=&-\frac{a^3 H_0^3 \xi ^3}{\mathcal{H}^3}\Bigg[  c_5 \xi ^2\l(3+3 \f{\ddot{\mathcal{H}}}{\mathcal{H}^3}+ 24 \f{\dot{\mathcal{H}}}{\mathcal{H}^2} -18\f{\dot{\mathcal{H}}^2}{\mathcal{H}^4}\r)\nn\\ 
&&-2 \xi c_4 \l(\f{\dot{\mathcal{H}}}{\mathcal{H}^2} -7\r)+2 c_3\Bigg]\,,\nn\\
\gamma_3 &=&-\frac{a^4}{\mathcal{H}^4} H_0^4 \xi ^4 \left(2 c_4 +3 c_5 \xi  \f{\dot{\mathcal{H}}}{\mathcal{H}^2}\right)\,,
\end{eqnarray}
\noindent
where dots stand for derivatives with respect to  conformal time, $\tau$. Finally, we have $2 \gamma_5 = - \gamma_4= \gamma_3 $ and $\gamma_6=0$.

We impose flat priors over the range $[0,10]$ for the models parameters $c_3$ and $\xi$ (when needed) and a flat prior over $[0,1]$eV for $\Sigma m_\nu$, when not set to zero.

Finally, the implementation of the CG models in EFTCAMB has been compared with other Einstein Boltzmann solvers for modified gravity in a recent work~\cite{Bellini:2017avd}, showing sub percent agreement at all scales of interest.
It demonstrates that the EFT approach is very robust to recover the linear perturbation theory from the covariant approach.

\subsection{Data sets}\label{Sec:Data}
 
 In the present analysis we consider the Planck measurements~\cite{Aghanim:2015xee,Ade:2015xua} of CMB temperature and polarization on large angular scales, limited to 
multipoles $\ell < 29$ (low-$\ell$ TEB likelihood) and the CMB temperature on smaller angular scales (PLIK TT likelihood, $30<\ell<2508$) along with Baryonic Acoustic Oscillations (BAO) 
measurements of the BOSS DR12 ({\it consensus} release)~\cite{Alam:2016hwk}. For the Planck likelihood, we also vary the nuisance parameters that are used to model foregrounds as well as instrumental and beam uncertainties. We shall refer to this data set as  PB ({\it Planck+BAO}).
We then complement it with results from local measurements of $H_0$~\cite{Riess:2016jrr}, weak gravitational lensing from the KiDS collaboration~\cite{Hildebrandt:2016iqg,Kuijken:2015vca,deJong:2012zb,Joudaki:2016kym} and the {\it Joint Light-curve Analysis} ``JLA'' Supernovae (SN) sample, as introduced in Ref.~\cite{Betoule:2014frx}. 
For the weak lensing data set, we decide to perform a cut at non-linear scales, since the predictions for CG models at those scales are not known very precisely. 
For this reason, we follow the analyses done in Refs.~\cite{Kitching:2014dtq,Ade:2015rim}, with the cut in the radial direction $k\le 1.5 \, h$Mpc$^{-1}$ and where the contribution from the $\xi^-$ correlation function is removed. In this way the analysis has been shown to be sensitive to the linear scales only (see Fig.~2 of~\cite{Ade:2015rim}).
We shall refer at this second data set as PBHWS, i.e. {\it Planck+BAO+$H_0$+WL+SN}.
This is the first time CG theories are being analyzed against such a wide data set, containing weak gravitational lensing data from KiDS.

\section{Results and Discussion}\label{results}

We analyze the three different CG models (i.e. $G_3$, $G_4$ and $G_5$) within four different cosmological scenarios, namely massless neutrinos and massive neutrinos with the three different hierarchies: normal, inverted and  degenerate. We will always report also the results for the fiducial $\Lambda$CDM cosmology, thus, we will analyze a total of $16$ different scenarios within the two data sets described in Section~\ref{Sec:Data}.
\begin{figure}
\begin{center}

\includegraphics[width=.49\textwidth]{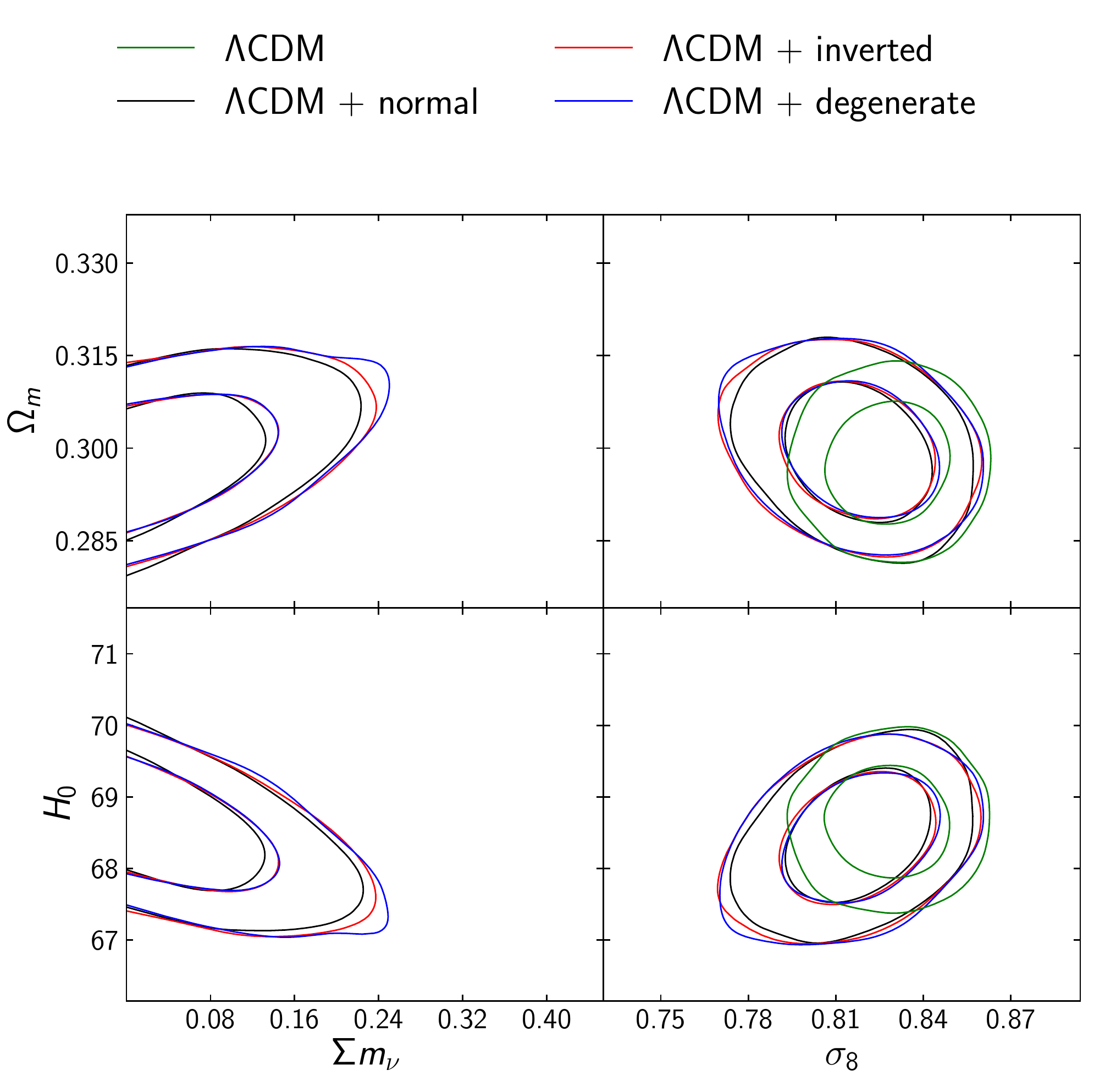}
\caption{The joint marginalized posterior of $\Lambda$CDM runs with   PBHWS  data set. The lines correspond to the 68\% C.L. and the 95\% C.L. regions. Different colours correspond to different neutrino scenarios as stated in the legend.
\label{fig:LCDM} 
}
\end{center}
\end{figure}

\begin{figure}
\begin{center}
\includegraphics[width=.49\textwidth]{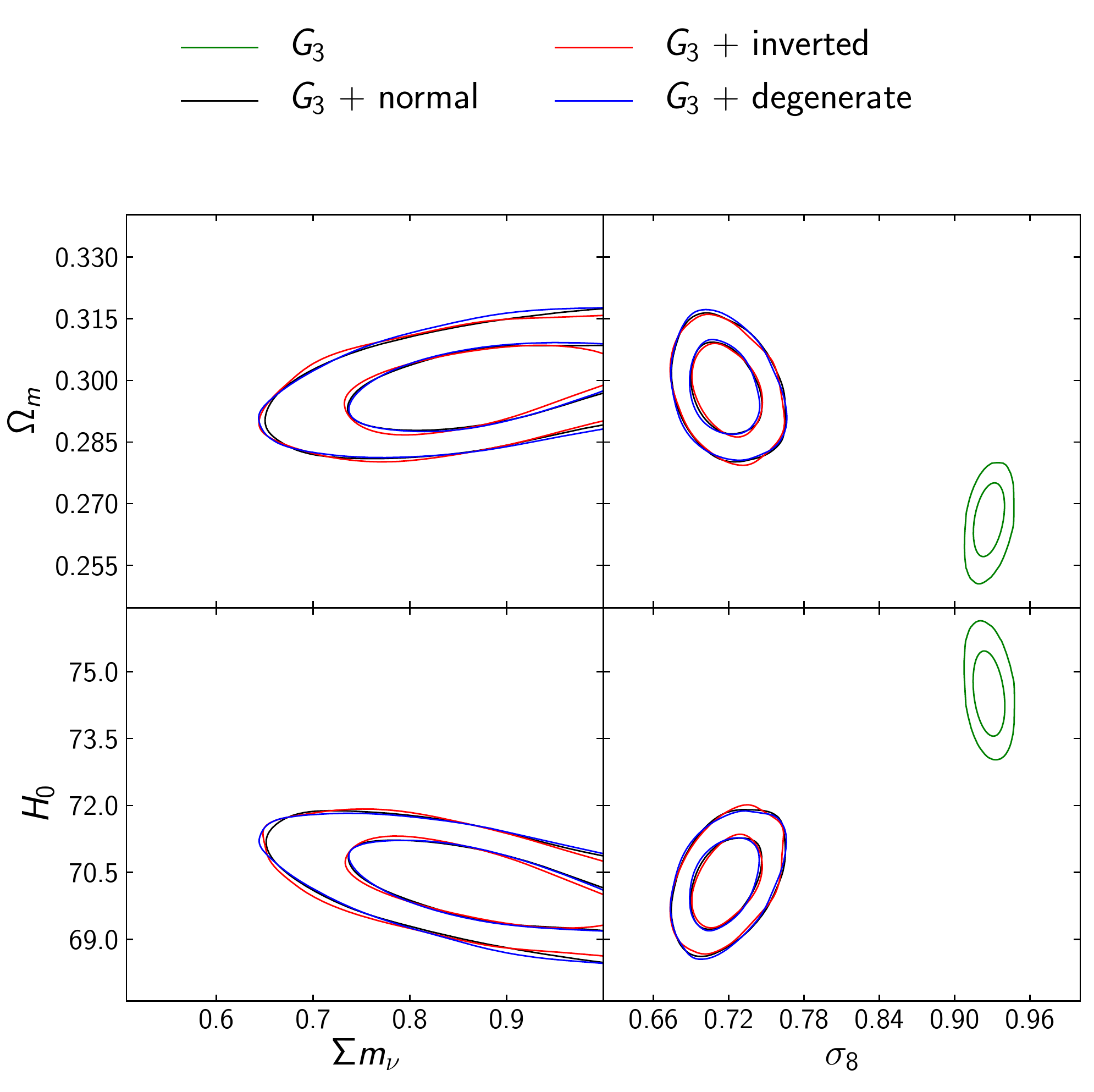}
\caption{The joint marginalized posterior of $G_3$ runs with the  PBHWS   data set. The lines correspond to the 68\% C.L. and the 95\% C.L. regions. Different colours correspond to different neutrino scenarios as stated in the legend.
\label{fig:G3} 
}
\end{center}
\end{figure}

\begin{figure}
\begin{center}
\includegraphics[width=.49\textwidth]{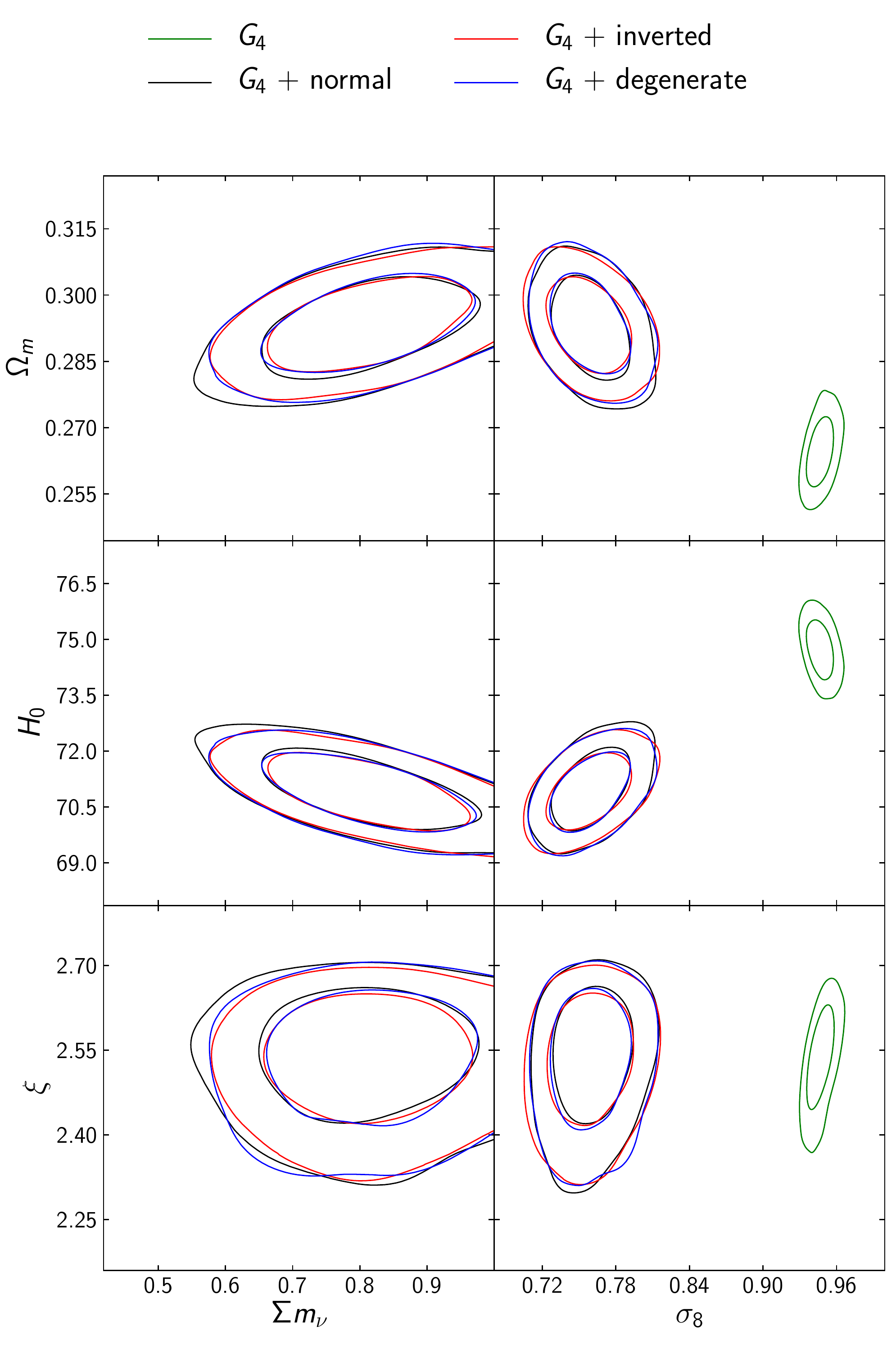}
\caption{The joint marginalized posterior of $G_4$ runs with the  PBHWS  data set. The lines correspond to the 68\% C.L. and the 95\% C.L. regions. Different colours correspond to different neutrino scenarios as stated in the legend.
\label{fig:G4} 
}
\end{center}
\end{figure}

\begin{figure}
\begin{center}
\includegraphics[width=.49\textwidth]{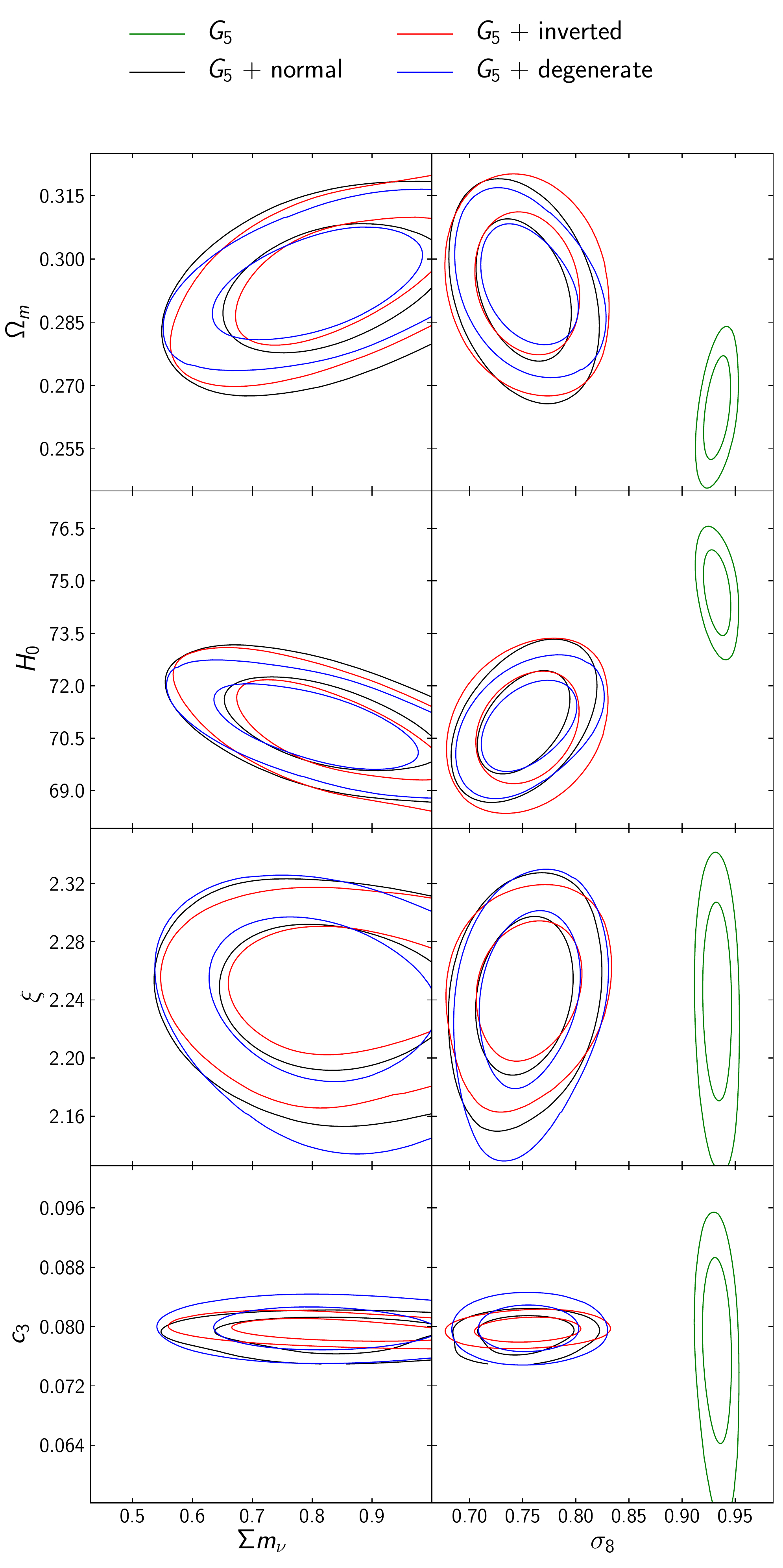}
\caption{The joint marginalized posterior of $G_5$ runs with the  PBHWS data set. The lines correspond to the 68\% C.L. and the 95\% C.L. regions. Different colours correspond to different neutrino scenarios as stated in the legend.
\label{fig:G5} 
}
\end{center}
\end{figure}

In Figures~\ref{fig:G3},~\ref{fig:G4} and~\ref{fig:G5}, we show the joint marginalized posterior distributions of the  cosmological parameters $\sigma_8$, $\Omega_m$, $H_0$, $\Sigma m_\nu$, along with the model parameters $\xi$ and $c_3$, obtained through the analysis of the $G_3$, $G_4$ and $G_5$ models for all the four neutrinos configurations. For comparison with the CG results, in Figure~\ref{fig:LCDM} we show the posterior distributions of the $\Lambda$CDM model. 

Let us first consider the case of zero neutrino mass and focus on the  effects that the scalar field of CG models has on the cosmological parameters.  The density parameter of baryons and cold dark matter, $\Omega_m$,  is shifted towards lower values in the CG cosmologies, on the contrary $\sigma_8$ increases.  From the different plots we can also see that the value of $H_0$ is enhanced easing the tension between CMB and the local measurements of $H_0$. Furthermore, we notice that there is no substantial impact of the different data sets on these parameters, then in the figures we show only the results obtained with the full  PBHWS data set. 

We then open the massive neutrino sector. As already noticed in Ref.~\cite{Barreira:2014jha}, all CG cosmologies are compatible with a detection of the neutrino mass ($\Sigma m_\nu \neq 0$). In general, the value of $\Sigma m_\nu$ for the CG cosmologies is higher  (a factor of $10$) if compared to the $\Lambda$CDM model best fit, see Table~\ref{tab_bestfit_params}. The cosmological parameters turn out to be affected by the presence of massive neutrinos, as expected, but the different hierarchies do not lead to a noticeable difference either on the cosmological or on the model parameters. The effect  of the massive neutrinos on $H_0$ is to contrast the impact of the scalar field; overall the value of $H_0$ in those runs remains higher compared to the $\Lambda$CDM one, but still compatible within $2\, \sigma$ error bars. $\Omega_m$ increases to higher values with respect to the zero neutrino mass cases, being now compatible with the  $\Lambda$CDM case. As a consequence, $\sigma_8$ assumes lower values with respect both to the zero neutrino mass CG cosmologies and the four $\Lambda$CDM scenarios. $\sigma_8$ assumes the lowest values for the $G_3$ case. The model parameter $\xi$  in $G_4$ and $G_5$ is not affected by the inclusion of massive neutrinos and, in general, in $G_5$ it assumes lower values. Finally, the values of $c_3$ in $G_5$, in all the four scenarios, are compatible within the errors, thus, the mean value of $c_3$ is not affected by the introduction of massive neutrinos, while its error bars are larger in the massless neutrinos scenario.

The impact of the two different data sets is to move to slightly bigger values $\Sigma m_\nu$ in the PBHWS  with respect to PB. This has the effect of reducing $\sigma_8$ from $0.933  \pm 0.006$ ($G_5$ with $\sum m_\nu =0$) to $0.75 \pm  0.02$ ($G_5$ with  $\sum m_\nu \neq 0$, independent of the hierarchy). For this reason, introducing massive neutrinos in CG cosmology has the effect of alleviating the CMB-WL tension~\cite{Barreira:2014jha}.

\begin{center}
\begin{table}
    \begin{tabular}{| l | c | c | c | c |}
    \hline
    Model &  $\hspace{0.1cm}$data set $\hspace{0.1cm}$ &  $\hspace{0.2cm}\chi^2\hspace{0.2cm}$  & $\hspace{0.2cm} \log_{10} B \hspace{0.2cm}$& $\hspace{0.1cm} \Delta \log_{10} B \hspace{0.1cm}$\\ \hline  \hline 
	$\Lambda$CDM		      &   PB  &     $5635$  & $-2459 $ & $0$\\
	$\Lambda$CDM+degenerate &   PB	 &     $5635$  & $-2459 $ & $0$\\
	$\Lambda$CDM+inverted      &	   PB  &	$5635$  & $-2459 $ & $0$\\
	$\Lambda$CDM+normal        &   PB	 &	$5635$  & $ -2459$ & $0$\\ \hline
	$G_3$			      &   PB	 &     $5674$ & $-2476 $& $-17$  \\
	$G_3$+degenerate      &	   PB  &	$5646$ & $ -2463$& $-4$  \\
	$G_3$+inverted	      &   PB  &	$5646$ & $ -2463$ & $-4$\\
	$G_3$+normal	      &   PB	 &     $5646$ & $-2464 $ & $-5$\\ \hline
	$G_4$			      &   PB	 &     $5667$ & $-2474 $ & $-15$\\
	$G_4$+degenerate      &	   PB	 &	$5643$ & $ -2464$& $-5$  \\
	$G_4$+inverted	      &   PB  &	$5644$ & $ -2463$& $-4$  \\
	$G_4$+normal	      &   PB	 &     $5645$ & $-2463 $& $-4$ \\ \hline
	$G_5$			      &   PB	 &     $5663$ & $-2473 $& $-14$  \\
	$G_5$+degenerate      &	   PB	 &	$5644$ & $-2465 $& $-6$ \\
	$G_5$+inverted	      &   PB  &	$5644$ & $-2465$& $-6$  \\
	$G_5$+normal	      &   PB  &     $5644$ & $ -2465$& $-6$ \\ \hline \hline
	$\Lambda$CDM	              & PBHWS  &       $6020$ & $-2628 $& $0$ \\
	$\Lambda$CDM+degenerate & PBHWS	&	$6020$ & $-2629 $& $0$ \\
	$\Lambda$CDM+inverted      &	 PBHWS 	&	$6020$ & $-2629$& $0$\\
	$\Lambda$CDM+normal        &  PBHWS	&	$6020$ & $-2629$& $0$\\ \hline
	$G_3$			    &   PBHWS	&      $6103$ & $ -2664$& $-36$\\
	$G_3$+degenerate    &	 PBHWS	&	$6052$ & $ -2640$& $-11$\\
	$G_3$+inverted	    &   PBHWS 	&	$6048$ & $ -2640$& $-11$\\
	$G_3$+normal	    &   PBHWS	&      $6047$ & $ -2640$& $-11$\\ \hline
	$G_4$			    &   PBHWS	&      $6078$ & $ -2652$& $-24$\\
	$G_4$+degenerate    &	 PBHWS	&	$6035$ & $ -2635$& $-6$\\
	$G_4$+inverted	    &   PBHWS	&	$6034$ & $-2635$& $-6$\\
	$G_4$+normal	    &   PBHWS	&      $6034$ & $ -2635$& $-6$\\ \hline
	$G_5$			    &   PBHWS	&      $6079$ & $ -2651$& $-23$ \\
	$G_5$+degenerate    &	 PBHWS	&	$6038$ & $ -2634$& $-5$\\
	$G_5$+inverted	    &   PBHWS 	&	$6036$ & $ -2634$& $-5$ \\
	$G_5$+normal	    &   PBHWS	&      $6038$ & $-2635 $& $-6$ \\
    \hline
     \end{tabular}
     \caption{\label{tab_bestfit} 
     Values of the best fit $\chi^2$ and of the Bayes factors ($\log_{10} B$) for the different CG models and data set combinations.  The Bayes factors difference is computed with respect to the $\Lambda$CDM model assuming the same neutrinos scenario and the same data set. Negative values of the Bayes factor disfavor the CG model.}   
 \end{table}
\end{center}

\begin{figure}
\begin{center}
\includegraphics[width=.49\textwidth]{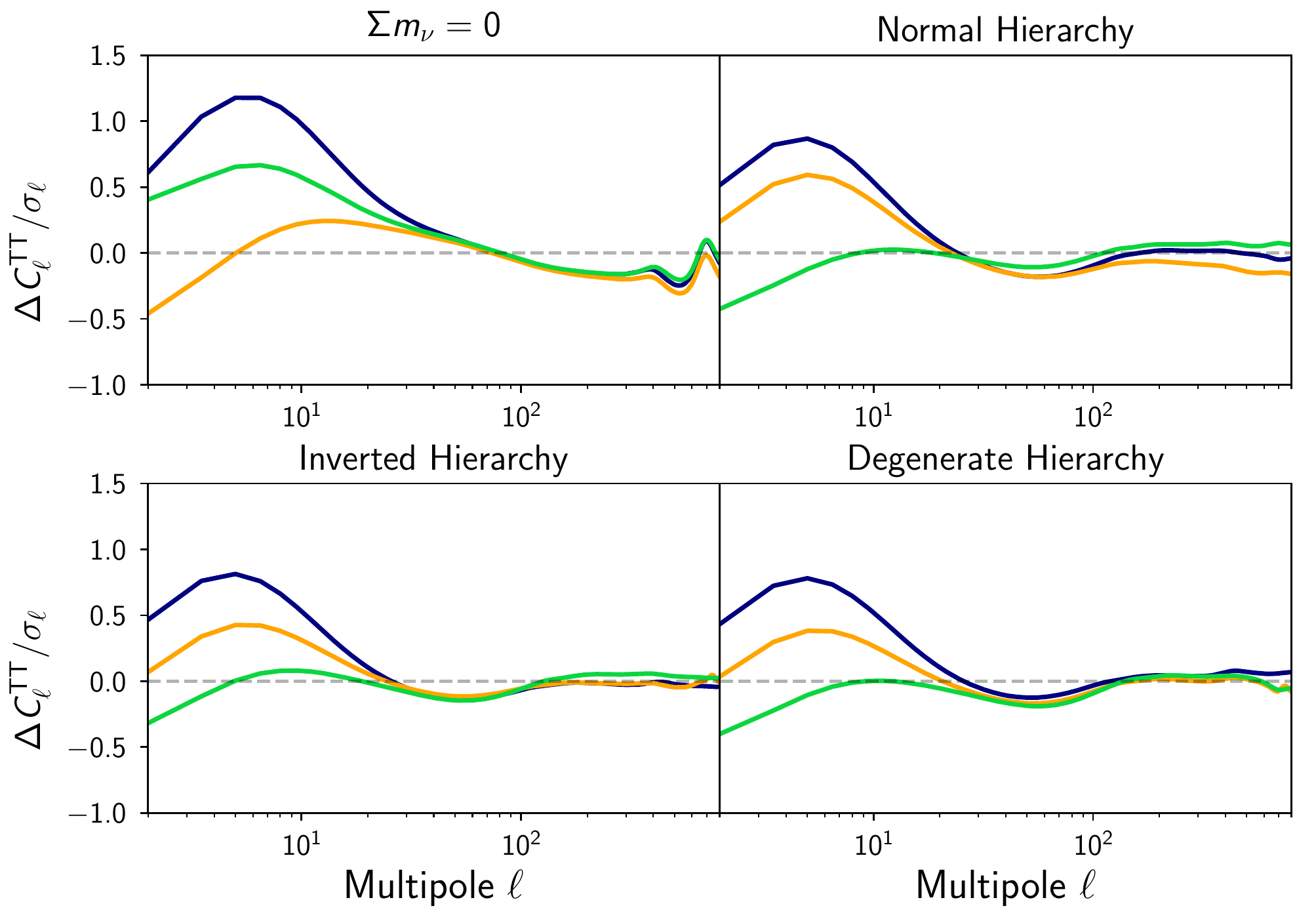}
\includegraphics[width=.49\textwidth]{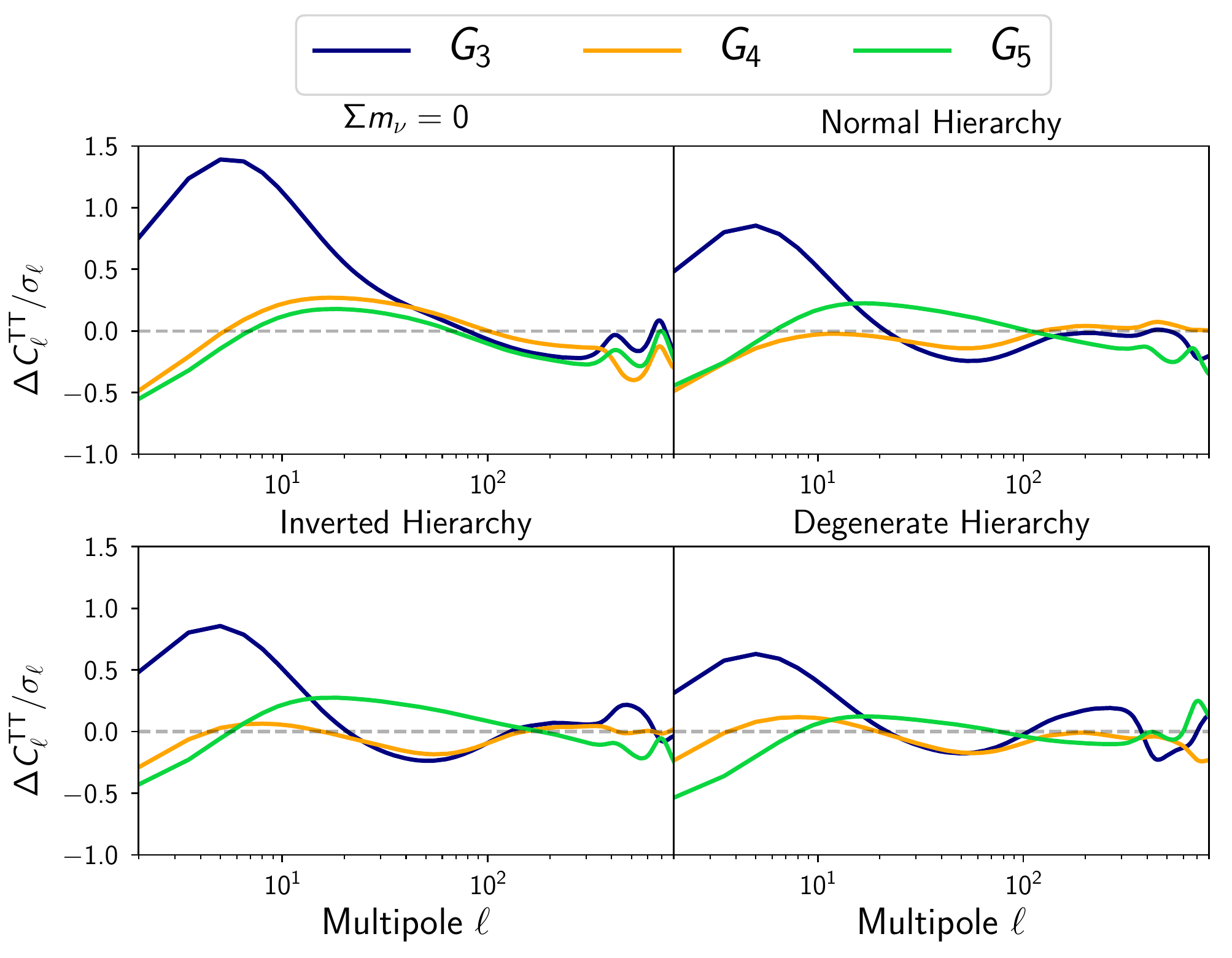}
\caption{Deviation in the CMB TT power spectra in units of TT variance, $\sigma_\ell = \sqrt{2/(2 \ell+1)} C_\ell^{\Lambda {\rm CDM}}$, for best fit parameters for PB (top) and PBHWS (bottom), computed with respect to $\Lambda$CDM.
\label{fig:BF} 
}
\end{center}
\end{figure}

\vspace{-0.8cm}
In Figure~\ref{fig:BF} we show the deviation of the best fit CMB TT power spectra, in unit of TT variance, $\sigma_\ell = \sqrt{2/(2 \ell+1)} C_\ell^{\Lambda {\rm CDM}}$, for each CG model, computed with respect to $\Lambda$CDM. In the upper panel we show the best fits from the analysis of PB, while in the lower one we show the results from PBHWS. As we can see, this deviation is larger in the case without massive neutrinos. We also see that there is almost no effect due to the hierarchy. Furthermore, we infer that $G_3$  always shows the worst fit to the CMB data, overestimating the $C_\ell^{\rm TT}$ at low $\ell$, while in $G_4$ and $G_5$ the presence of extra parameters allows for a better fit. Nevertheless, when passing from $G_4$ to $G_5$, and thus allowing for a new free parameter (i.e. $c_3$), the fit does not look improved: in fact, we cannot see substantial difference between the two cases, especially when looking at the full PBHWS data sets results.
We find that the choice of the hierarchy does not leave any signatures on the best fit of the matter power spectrum, as it is possible to grasp from the mean values obtained for $\sigma_8$ in the different cases.

Finally, we perform a complete statistical analysis, including the best chi squared and Bayesian evidence. In Table~\ref{tab_bestfit}, we show the values of the best fit $\chi^2$ for the different runs and  the Bayesian evidence factors ($\log_{10} B$), computed as defined in~\cite{Heavens:2017afc, DeBernardis:2009di}. The last column is the difference between the Bayes factor of the i-th CG model and the value obtained in the $\Lambda$CDM run, with same hierarchy and data set. This value is interpreted following the Jeffreys' scale that judges odds in favor of one model exceeding $100:1$, or $\Delta \log_{10}B > 2$, to be decisive in favor of the model.
Looking at the $G_3$ best fits, we see that the $\Delta \log_{10} B$ decreases drastically in the runs without massive neutrinos. The situation gets better with massive neutrinos, but without showing any preference for a hierarchy.  The same trend can be also noticed by looking at the right column in Table~\ref{tab_bestfit}. The $\Delta \log_{10} B<-4$ and the situation gets worst when the complete data set (PBHWS) is used ($\Delta \log_{10} B<-11$). This means that, for both data sets, $G_3$ is a worst fit to the data, compared with $\Lambda$CDM. A similar result has been shown in Ref.~\cite{Renk:2017rzu}, where the authors find that $G_3$ is effectively ruled out when constrained against ISW data. Compared to such analysis, we use a different data set, which includes weak lensing, obtaining similar results.  
One would expect this situation to improve significantly in the $G_4$ and $G_5$ runs, since the presence of more parameters ($\xi$ for $G_4$, $\xi$ and $c_3$ for $G_5$) should give to the model more freedom to adapt to the data. However, these are not the cases and the differences in the $\log_{10} B$ exceed the $4$ units, for both models with massive neutrinos regardless of the data set considered. Such differences in Bayes factor exceed the $20$ units in the scenarios with zero neutrino mass in the runs with the full data set.  
In summary, the higher Bayesian evidence is always found when fitting with $\Lambda$CDM, with no visible improvement when allowing the neutrino mass to vary and no effect coming from the hierarchy, regardless of the data set combination. Thus, the  negative values of $\Delta \log_{10} B$ indicate that all the different CG models are strongly disfavored with respect to $\Lambda$CDM. Because such  discrepancies in the $\log_{10} B$ comparison are very large, we consider our conclusion exhaustive and very robust.

In conclusion, we claim that all CG models are statistical ruled out by cosmological data. Thus, we confirm that the G3 model is excluded by data as previously noticed in ref.~\cite{Renk:2017rzu}. More remarkably, the constraining power of the data sets used in this analysis allows us to exclude G4 and G5 as well, for the first time only by means of cosmological data. Interestingly this latter result is in line with the theoretical implications of the measurements  of GW170817 and its electromagnetic counterpart which severely  constrain both G4 and G5~\cite{Creminelli:2017sry,Ezquiaga:2017ekz,2017arXiv171006394B,Sakstein:2017xjx, Bettoni:2016mij}. 

\begin{center}
\begin{table*}
    \begin{tabular}{| l | c | c | c | c | c | c |}
    \hline
    Model &  $\sigma_8$ &  $\Omega_m$   & $H_0$ & $\Sigma m_\nu$ & $\xi$ & $c_3$\\ \hline  \hline 
	$\Lambda$CDM	              & $ 0.83  \pm 0.01$ & $0.298 \pm 0.007 $ & $ 68.7  \pm  0.5 $ & -&-&-     \\
	$\Lambda$CDM+degenerate & $ 0.82  \pm 0.02$ & $ 0.300 \pm 0.007 $ & $ 68.4 \pm  0.6 $ & $0.08  \pm 0.06 $ & - & - \\
	$\Lambda$CDM+inverted      &	 $ 0.82  \pm 0.02$ & $0.300 \pm 0.007 $ & $ 68.4  \pm  0.6 $ & $0.07  \pm 0.06 $ & - & -\\
	$\Lambda$CDM+normal        &  $0.82   \pm 0.02$ & $0.300 \pm 0.007 $ & $ 68.4  \pm  0.6 $ & $0.07  \pm 0.06 $ &- &- \\ \hline
	$G_3$			    &  $0.928  \pm 0.007$ & $ 0.266 \pm 0.004 $ & $74.6  \pm  0.4 $ & - & - & -  \\
	$G_3$+degenerate    &	$0.72  \pm 0.02$ &     $ 0.298 \pm 0.007 $ & $70.3  \pm  0.6 $ &  $0.85   \pm 0.08 $ & - & -  \\
	$G_3$+inverted	    &  $0.72  \pm 0.02$ &     $ 0.297 \pm 0.007 $ & $70.3  \pm  0.6 $ &  $0.85  \pm 0.08 $ & - & -\\
	$G_3$+normal	    &  $0.72  \pm 0.02$ &     $ 0.298 \pm 0.007 $ & $70.3   \pm  0.6 $ & $0.85   \pm 0.08 $ & - & -    \\ \hline
	$G_4$			    &  $0.948  \pm 0.008$ & $ 0.264 \pm 0.005 $ & $74.7  \pm  0.5 $ & - 				 & $ 2.53 \pm 0.06 $ & -   \\
	$G_4$+degenerate    &	$0.76  \pm 0.02$ &     $ 0.293 \pm 0.007 $ & $70.9  \pm  0.6 $ & $0.80  \pm 0.09 $ & $ 2.53 \pm  0.07$ & -\\
	$G_4$+inverted	    &  $0.76  \pm 0.02$ &     $ 0.293 \pm 0.007 $ & $70.9  \pm  0.6 $ & $0.80   \pm 0.09 $   & $ 2.53 \pm  0.07 $ & - \\
	$G_4$+normal	    &  $0.76  \pm 0.02$ &     $ 0.293 \pm 0.007 $ & $71.0  \pm  0.7 $ & $0.80  \pm 0.09 $    & $ 2.53 \pm  0.08 $ & - \\ \hline
	$G_5$			    &  $0.933  \pm 0.006$ & $ 0.264 \pm 0.005 $ & $74.7  \pm  0.5 $ & -                               & $2.23 \pm  0.03 $ & $0.076 \pm  0.006$  \\
	$G_5$+degenerate    &	$0.76 \pm  0.02 $ &     $ 0.294 \pm  0.006 $ & $ 70.9 \pm  0.6 $ & $ 0.80 \pm  0.09 $ & $2.24 \pm  0.03 $ & $0.080 \pm  0.001 $  \\
	$G_5$+inverted	    &  $0.75 \pm  0.02 $ &     $ 0.294 \pm  0.007 $ & $ 70.9 \pm  0.7 $ & $ 0.81 \pm  0.09 $ & $2.24 \pm  0.02 $ & $0.0796 \pm  0.0007$  \\
	$G_5$+normal	    &  $0.75 \pm  0.02 $ &     $ 0.292 \pm  0.007 $ & $ 71.0 \pm  0.6 $ & $ 0.81 \pm  0.09 $ & $2.24 \pm  0.02  $ & $0.079 \pm  0.001$ \\
    \hline
     \end{tabular}
     \caption{\label{tab_bestfit_params} Constraints on cosmological and model parameters at 1$\sigma$. These values are obtained through the analysis of the full  PBHWS data set. }  
 \end{table*}
\end{center}

\section{Conclusion}\label{conclusion}

In this work we have explored the phenomenology of Covariant Galileons in light of the latest releases of cosmological data. For the first time in literature CG have been constrained against a wide and comprehensive dataset, containing WL measurements from the KiDS collaboration. As an additional degree of freedom of the theory, we have allowed for three different mass hierarchies and we have investigated the corresponding bounds on the CGs and cosmological parameters with current data.
We have first considered three different CG classes, Cubic ($G_3$), Quartic ($G_4$) and Quintic ($G_5$) Galileon in order to distinguish the effect of the different terms in the Lagrangian. Then for each of them, we have considered four different scenarios: a cosmology with $\Sigma m_\nu = 0$, and cosmologies with three different mass hierarchies, i.e. normal, inverted and degenerate. We have included for the same scenarios the $\Lambda$CDM model for comparison and actually distinguish the impact of the additional scalar field and that  of massive neutrinos.

For the analysis presented in this work, we used two separate data sets as explained in Section~\ref{method}, but we did not find any significant improvement when comparing the results from PB ({\it Planck+BAO}) with the ones from the complete PBHWS ({\it Planck+BAO+$H_0$+Weak Lensing+Supernovae}) data set. Thus, we showed the results of the complete data set containing for the first time measurements from the weak gravitational lensing survey KiDS.
   
We confirm that a CG cosmology implies a non zero neutrino mass, with $\Sigma m_\nu = 0 $ giving always a bad fit to data in both $G_3$, $G_4$ and $G_5$, see Table~\ref{tab_bestfit} and Figure~\ref{fig:BF}. The value of the neutrino mass is higher than in $\Lambda$CDM: for the normal hierarchy we got $0.85 \pm 0.08$ eV for $G_3$, $0.80 \pm 0.09$ eV for $G_4$ and  for $0.81 \pm 0.09$ eV $G_5$, while $\Lambda$CDM gives $0.07 \pm 0.06$ eV. In view of this relaxed bound, we find not sizeable difference with the other mass hierarchies, indicating that current data cannot pick up the subtle features in the matter power spectrum due to the differences in the relativistic to non-relativistic transition redshifts. 
When including massive neutrinos, the models considered seem to be really efficient in solving the CMB-low $z$ tensions, by preferring an higher value of $H_0$ and lowering $\sigma_8$ (see Table~\ref{tab_bestfit_params}).  

Nevertheless, a careful statistical analysis, based on the $\chi^2$ and the Bayesian Evidence comparison, shows that all the CG models are a much worse fit to the data, compared to $\Lambda$CDM, whatever hierarchy is considered (see Table~\ref{tab_bestfit}) even the models, like G4 and G5, that have extra free parameters. The results of such simple analysis are strong enough to confidently rule out all the CG models.
In the case of $G_3$, this was already noticed in~\cite{Renk:2017rzu}, where the authors find that the Cubic Galileon is effectively ruled out by ISW data.
From Figure~\ref{fig:BF} we can see that, whatever CG configuration or hierarchy used, the model always give a bad fit of the CMB TT power spectrum at low~$\ell$. We can see that including $\sum m_\nu \neq 0$ helps in lowering the $\chi^2$ by a factor of $3$, but still the $\Delta \log_{10} B$, computed with respect to $\Lambda$CDM, is large.
These results allow us to claim for the first time that the entire class of CG models are statistically ruled out by cosmological data only. 

Recently, in~\cite{Creminelli:2017sry,Ezquiaga:2017ekz,2017arXiv171006394B,Sakstein:2017xjx, Bettoni:2016mij}, it was shown that the  measurements of the electromagnetic counterpart of the gravitational wave GW170817~\cite{GBM:2017lvd,Coulter:2017wya} set stringent theoretical constraints on the Quartic and Quintic Lagrangians, practically ruling out their contribution from the action, unless they reduce to a standard conformal coupling. The Cubic Galileon is not affected by these bounds. We have shown that cosmological data alone, are able to exclude the viability of all CG models.

\vspace{0.1cm}


\begin{acknowledgments}
We thank A. Barreira and B. Li for comparing our results at the early stages of this work and M. Zumalacarregui for useful comments.
The research of NF is supported by Funda\c{c}\~{a}o para a  Ci\^{e}ncia e a Tecnologia (FCT) through national funds  (UID/FIS/04434/2013) and by FEDER through COMPETE2020  (POCI-01-0145-FEDER-007672).
AS and SP acknowledge support from the NWO and the Dutch Ministry of Education, Culture and Science (OCW), and also from the D-ITP consortium, a program of the NWO that is funded by the OCW.
NF, SP and AS acknowledge the COST Action  (CANTATA/CA15117), supported by COST (European Cooperation in  Science and Technology).
BH is supported by the Beijing Normal University Grant under the reference No. 312232102, National Natural Science
Foundation of China Grants No. 210100088, No. 210100086. 
BH is also partially supported by the Chinese National Youth Thousand Talents Program and the Fundamental Research Funds for
the Central Universities under the reference No. 310421107.
MR is supported by U.S. Dept. of Energy contract DE-FG02-13ER41958.
\end{acknowledgments}

\appendix


\bibliographystyle{aipnum4-1}
\bibliography{galileons}

\end{document}